\documentstyle[12pt,psfig,aaspp4]{article}	

\def\gradip{\hbox{\rlap{\hbox{.}}\raise 5.truept \hbox{{\small $\circ$}}}}

\catcode`\@=11
\def\gsim{\ifmmode{\mathrel{\mathpalette\@versim>}}
    \else{$\mathrel{\mathpalette\@versim>}$}\fi}
\def\lsim{\ifmmode{\mathrel{\mathpalette\@versim<}}
    \else{$\mathrel{\mathpalette\@versim<}$}\fi}
\def\@versim#1#2{\lower 2.9truept \vbox{\baselineskip 0pt \lineskip
    0.5truept \ialign{$\m@th#1\hfil##\hfil$\crcr#2\crcr\sim\crcr}}}
\catcode`\@=12

\begin{document}

\title{Testing intermediate-age stellar evolution models with VLT photometry
of LMC clusters. I. The data\altaffilmark{1}}

\author{
Carme Gallart\altaffilmark{2,3},
Manuela Zoccali\altaffilmark{5},
Giampaolo Bertelli\altaffilmark{6,9},
Cesare Chiosi\altaffilmark{7},
Pierre Demarque\altaffilmark{4},
Leo Girardi\altaffilmark{8},
Emma Nasi\altaffilmark{9},
Jong-Hak Woo\altaffilmark{4},
Sukyoung Yi\altaffilmark{10}
}

\altaffiltext{1}{Based on observations collected at the European Southern
Observatory, Paranal, Chile (ESO N$^o$ 64.L-0385)}

\altaffiltext{2}{Andes Prize Fellow, Universidad de Chile and Yale University }
\altaffiltext{3}{Currently: Ram\'on y Cajal Fellow. Instituto de Astrof\'\i sica de 
	Canarias, 38200 Tenerife, Canary Islands, Spain; carme@iac.es}
\altaffiltext{4}{Department of Astronomy, Yale University, P.O. Box 208101, New Haven, 
	CT 06520-8101; demarque@astro.yale.edu, jhwoo@astro.yale.edu}
\altaffiltext{5}{European Southern Observatory, Karl Schwarzschild Strasse 2,
	D-85748 Garching bei M\"unchen, Germany; mzoccali@eso.org}
\altaffiltext{6}{National Council of Research, IAS-CNR, Rome, Italy; 
	bertelli@pd.astro.it}
\altaffiltext{7}{Dipartimento di Astronomia dell'Universit\`a di Padova, Vicolo 
	dell'Osservatorio 5, I-35122 Padova, Italy; chiosi@pd.astro.it}
\altaffiltext{8}{Osservatorio Astronomico di Trieste, Via Tiepolo 11, I-34131 Trieste, 
	Italy girardi@pd.astro.it}
\altaffiltext{9}{Osservatorio Astronomico di Padova, Vicolo dell'Osservatorio 5, 
	35122 Padova, Italy; nasi@pd.astro.it}
\altaffiltext{10}{Astrophysics, University of Oxford, Keble Road, Oxford, OX1 3RH, 
	United Kingdom; yi@astro.ox.ac.uk}

\begin{abstract}

This is the first of a series of three papers devoted to the
calibration of a few parameters of crucial importance in the modeling
of the evolution of intermediate-mass stars, with special attention to
the amount of convective core overshoot. To this end we acquired deep
$V$ and $R$ photometry for three globular clusters of the Large
Magellanic Cloud (LMC), namely NGC~2173, SL~556 and NGC~2155, in the
age interval 1--3 Gyr.  In this first paper, we describe the aim of
the project, the VLT observations and data reduction, and we make
preliminary comparisons of the color-magnitude diagrams with both
Padova and Yonsei-Yale isochrones. Two following papers in this series
present the results of a detailed analysis of these data,
independently carried out by members of the Yale and Padova stellar
evolution groups. This allows us to compare both sets of models and
discuss their main differences, as well as the systematic effects that
they would have to the determination of the ages and metallicities of
intermediate-age single stellar populations.

\end{abstract}


\section{INTRODUCTION}
\label{intro}

In an epoch of extraordinary discoveries on the high-z Universe, we
still have gaps in our understanding of stellar evolution, which is in
fact essential for the correct interpretation of the light of
distant galaxies. Galactic star clusters have traditionally provided a
major way to study stellar evolution.  However, our Galaxy only
contains stars in age and chemical composition domains which reflect
its particular history, and these are the domains which have been
relatively well explored.  Extrapolation outside the limits of the
well explored age and metallicity range is not always safe. For
instance, it has become increasingly apparent in recent years that
metal content can affect stellar evolution in unexpected ways. The UV
upturn in elliptical galaxies is a case in point.  Studies of very
metal rich stellar systems revealed that, contrary to simple
expectations, old metal rich stellar populations do not simply become
redder in all wavebands as they evolve, but rather produce a
population of UV bright stars (Greggio \& Renzini, 1990; Horch,
Demarque \& Pinsonneault, 1992; Fagotto et al. 1994; Yi, Demarque \&
Oemler 1998). On the other hand, understanding the evolution of
extremely metal-poor and metal-free (Population III) stars will
similarly be essential for the interpretation of primordial, high-z
stellar populations.

In intermediate-age stellar populations, the color-magnitude diagram
(CMD) and the luminosity function are affected by convective core
overshoot.  While there is not a general agreement on the efficiency
of this process, and its extent is not well established, most
researchers agree that this parameter affects significantly the
morphology of the CMD of these clusters and naturally the
determination of their ages as well (e.g. Rosvick \& VandenBerg 1998;
Keller, Da Costa \& Bessell 2001; Meynet, Mermilliod \& Maeder 1993;
Carraro et al. 1993; Demarque, Sarajedini \& Guo 1994).  This
uncertainty is thus also a problem for the spectral dating of distant
stellar systems from their integrated light (Heap et al.  1998; Yi et
al. 2000).  In addition, deep CMDs of intermediate-age clusters are
essential to help disentangle the relative importance of other poorly
determined parameters, such as mass loss during red giant branch (RGB)
evolution, internal rotation (important for the more massive objects),
and helium content. In particular, the adopted value of the parameter
$\delta Y/\delta Z$ of helium enrichment significantly affects the
mass luminosity relation of stellar models, and consequently their
evolutionary lifetimes.

The papers in this series deal mainly with the problem of convective core
overshoot in intermediate-age stellar populations, and therefore we
will briefly review some of the recent work in the subject. Early
observational arguments in favor of convective core overshoot come from
clusters like Pleiades (Maeder \& Mermilliod 1981; Mazzei \& Pigatto
1989) and from other Galactic clusters, as discussed in great detail
by Maeder \& Mermilliod (1981) and Mermilliod \& Maeder (1986).
Barbaro \& Pigatto (1984) and Chiosi \& Pigatto (1986) argued for
overshoot in stars with masses in the range $1.5 - 2.2 M_{\odot}$ by
pointing out that the base of the RGB is not well
populated in clusters with age 1 -- 2 Gyr (whereas it is in
older clusters), as if degenerate He-ignition and He-flash were avoided
for this mass range, in contrast with classical models.  More recent
studies include those of Aparicio et al. (1990), Carraro et
al. (1993), Meynet et al. (1993); Rosvick \& VandenBerg (1998);
Demarque et al. (1994); Dinescu et al. (1995); Kozhurina-Platais et
al. (1997) which deal with Galactic open clusters of ages 1.0 to 6.0
Gyr. All conclude that a certain amount of convective core overshoot is
preferred to reproduce the CMDs. Being Galactic, all these clusters
have metallicity close to solar, and therefore do not allow the
possibility of testing the dependence of convective core overshoot on
metallicity.

The Magellanic Clouds (MC) offer an unusual opportunity to test
stellar evolution since populous clusters of a wide variety of ages
and chemical compositions can be observed and compared at effectively
the same distance. In particular, they contain young and
intermediate-age metal-poor stellar populations that are absent in the
Galaxy.  A number of young MC clusters (age $\le$ 500 Myr) have been
used by Lattanzio et al (1991), Vallenari et al. (1991, 1994),
Stothers \& Chin (1992), Brocato, Castellani \& Piersimoni (1994),
Chiosi et al. (1995), Testa et al. (1999) and Barmina et al. (2002) to
constrain evolutionary models of young stars. It is interesting to
note that these different studies do not always agree on the preferred
amount of convective core overshoot to best reproduce the observed
CMDs.  In particular there is a long lasting debate about the young
LMC cluster NGC 1866, where turnoff stars have about $4 - 5
M_{\odot}$: Chiosi et al. (1989a,b) and Brocato \& Castellani (1988),
Lattanzio et al. (1991) and Brocato et al. (1994), Testa et al. (1999)
and Barmina et al. (2002), are pairs of papers presenting
systematically opposite conclusions (for or against overshooting).  To
briefly summarize only the most recent ones, Testa et al. (1999)
obtained deep photometry ($V<24$) of a wide region of this cluster,
and concluded that the best fit to both the integrated luminosity
function and the magnitude of the horizontal branch is achieved by
classical models, without overshooting, provided that a 30$\%$
fraction of binaries is allowed. However, a new analysis of the same
data with a more accurate completeness correction and normalization,
by Barmina et al. (2002), showed that models with overshoot give a
better fit to both the overall morphology of the CMD and to the
integrated luminosity function of main sequence stars, reproducing the
correct ratio of main sequence to post-main sequence stars.

Thanks to the brightness of the stars in the young MC clusters, the
above mentioned observations were feasible using medium size
telescopes. Instead, our aim is to explore the less studied regime of
intermediate-age, low metallicity clusters, which turnoff is by
definition fainter, and therefore only 8-meter size telescopes under
excellent seeing conditions, or the HST, can give the necessary high
quality photometry.  Our intention is also to independently analyze
the data with two of the most recent sets of stellar evolutionary
models widely in use, namely, those of Padova (Girardi et al. 2000)
and Yonsei-Yale (Yi et al. 2001, $Y^2$ thereafter).

In this paper, we discuss the project setup (Sec. 2), and we present
the data obtained with VLT: the observations, photometry and crowding
tests are presented in Sec. 3, while the procedure used to
statistically substract the LMC field stars from the cluster CMDs is
discussed in Sec. 4. In Sec. 5, we briefly discuss the main
differences between the stellar evolution models of Girardi et
al. (2000) and Yi et al. (2001). In Sec. 6, a preliminary comparison
of the cluster CMDs with these stellar evolutionary models is
performed. In two forthcoming papers that follow in this same issue,
these data are independently analyzed in detail with both the $Y^2$
(Woo et al. 2002) and Padova (Bertelli et al. 2002) models, in such a
way that some feedback on the stellar evolution models used can be
provided.

\section{OBSERVATIONAL GOAL AND TARGET SELECTION}
\label{goal}

The goal of this project was to obtain accurate $V,R$ photometry down
to about 4 magnitudes below the main sequence turnoff of a number of
intermediate-age MC clusters, in order to construct CMDs and
luminosity functions to be used as test for stellar evolutionary
models computed independently by members of the Yale Group (Demarque,
Yi, Woo)and the Padova Group (Bertelli, Chiosi, Girardi, Nasi).

We originally proposed to observe eight clusters, in pairs of 1, 2, 3
and 5 Gyr each (according to the ages quoted in the literature) to be
able to check the extent of the fluctuations in the properties of
clusters (e.g. luminosity function) of the same age. The candidates
were selected according to: {\it a)} their age (from Geisler et
al. 1997; Sarajedini 1998; Mighell, Sarajedini \& French 1998), {\it
b)} their position in the galaxy, trying to avoid areas densely
populated by field stars, and {\it c)} their richness. The original
list of clusters included NGC~2209 and NGC~2249 ($\simeq$ 1 Gyr old),
NGC~1651 and NGC~2162 ($\simeq$ 2 Gyr old), NGC~2173 and SL~556
($\simeq$ 3 Gyr old) and NGC~2155 and Kron~3 ($\simeq$ 5 Gyr old). The
first four younger clusters were to be observed with NTT down to a
limiting magnitude of V=23.5 and 24.5, while the rest were proposed
for FORS1@VLT, with the goal of reaching limiting magnitudes V=25.5
(for NGC~2173 and SL~556), V=26.0 for NGC~2155 and V=26.3 for Kron~3.
Due to the high resolution needed to obtain accurate photometry in
such rich and distant clusters, we requested service mode observations
imposing a seeing constraint of FWHM $\le$0.6\arcsec. In the end, due
to observational problems and restrictions, no observations with NTT
were performed for this program, while 3 out of the 4 clusters to be
observed with VLT were indeed observed, namely, NGC~2155, NGC~2173 and
SL~556.

In what follows we summarize the results of previous studies of these
clusters, concentrating in particular on the metallicity
determinations, being this an important input parameter for our
analysis. The distance modulus of the LMC, being the first step of the
distance scale ladders, has been the subject of an intense debate in
recent years and trying to summarize it goes beyond the scope of this
paper: the generally accepted value of $(m-M)_V=18.5$ (Benedict et
al. 2002) has been adopted, allowing a variation of $\pm 0.2$ in the
isochrone fit, which can be justified by volume effect of the LMC on
the distance modulus of each cluster. Small variation of the reddening
around $E(B-V)\approx0.04$ along the different lines of sight of the
three clusters, will be allowed in our analysis as well.

Available determinations of the metallicity of these clusters are very
inhomogeneous, first of all because of the different methods applied,
and second because some of them lead to the determination of the
global metallicity, [M/H]=$log Z/Z_{\odot}-logX/X_{\odot}$ (e.g. when
using isochrone fitting), while some others give a measure of the iron
abundance, [Fe/H]=$log Fe/Fe_{\odot}-logX/X_{\odot}$
(e.g. spectroscopic or photometric methods calibrated against
[Fe/H]). Obviously the two quantities are different if the
alpha-element ratio of the cluster stars is different from solar, as
it is for galactic globular clusters.  Since there currently is no
high dispersion spectral analysis of the MC clusters from which one
could infer their alpha-element enhancement, in what follows we will
keep the distinction among the two chemical abundances, as measured by
the various authors, reminding the reader that {\it if} the
alpha-element enhancement of these LMC clusters were the same as that
of galactic globular clusters (i.e., [$\alpha$/Fe]$\sim$ 0.3, Carney
1996) then their global metal abundance [M/H] can be assumed to be
$\sim 0.21$ dex higher than the iron content [Fe/H] (Salaris, Chieffi
\& Straniero 1993).

Low resolution spectra of four stars in {\bf NGC~2173} were obtained
by Cohen (1982), who found a global metallicity of [M/H]=--0.75. A
lower value ([Fe/H]=--1.4) was obtained by Bica, Dottori and Pastoriza
(1986) with integrated photometry, while the Ca triplet analysis of
two stars, by Olszewski et al. (1991), gave [Fe/H]=--0.24. $B,V$
photometry for this cluster has been published by Mould, Da Costa \&
Wieland (1986), who estimate [M/H]=--0.8 from the RGB slope. They
found an age of either 1.4 or 2 Gyr, when adopting a long (18.7) or
short (18.2) absolute distance modulus, respectively. More recent
photometry for NGC~2173 has been published by Corsi et al. (1994), who
were looking for global properties of intermediate-age populations and
therefore do not analyze the CMD of this cluster individually, but
only superimposed to that of 10 other clusters. Finally, Geisler et
al. (1997), adopting the metallicity [Fe/H]=--0.24, as found by
Olszewski et al. (1991), estimate a cluster age of 2 Gyr, in agreement
with that found by Mould et al. (1986).

{\bf SL~556} (Hodge~4) is a relatively less studied cluster. No
spectroscopic studies have been carried on so far. $U,B,V$ photometry
has been published by Mateo \& Hodge (1986) who estimate a metallicity
of [Fe/H]=--0.7 using various photometric indexes. Olzewski et al. (1991)
determined [Fe/H]=--0.15, while more recent photometric studies have
been published by Sarajedini (1998) and Rich et al. (2001), who analyzed
the same HST data, and, adopting a metallicity [M/H]=--0.68 both found
an age between 2 and 2.5 Gyr.

{\bf NGC~2155} has been studied by Bica et al. (1986)
who performed DDO integrated photometry of a number of MC clusters.
They estimated a metallicity of [Fe/H]=--1.2. More recently, a
significantly higher value ([Fe/H]=--0.55) has been found by Olszewski
et al. (1991) who performed Ca II triplet spectroscopy of three cluster
giants. Although possibly more accurate than the previous integrated
photometry, this method is not very accurate because its calibration
in the metal-rich ([Fe/H]$>-1$), relatively young populations is more
uncertain than for the old, metal poor ones (Pont et al. 2002).
Indeed, for the three clusters studied here, the metallicities quoted
in Olszewski et al. (1991) are systematically higher (by $\sim 0.5$)
than any other published value.  A more recent determination of the
cluster metallicity from the RGB slope comes from the HST photometry
published by Sarajedini (1998).  The latter estimates [Fe/H]=--1.1, and
an age of $\sim 4$ Gyr.  The same HST data used by Sarajedini have
been re-analyzed by Rich et al. (2001) and compared with the
isochrones by Girardi et al. (2000), giving a global metallicity of
[M/H]=--0.68 and an age of 3.2 Gyr. 


\section{THE DATA}
\label{sec_obs}

The three LMC clusters  NGC~2173, SL~556 and NGC~2155 were observed
with FORS1@VLT-UT1, in service mode, through the filters $V$ and $R$.
Figure~\ref{frame} shows a median of all the frames for SL~556, as an
example of the image quality of the data. FORS1 was used in normal
resolution: its pixel size of 0.2 $''$/px allows a good PSF sampling,
while the relatively large field of view ($6.8\times6.8$ arcmin)
allowed us to sample, in the field outermost regions, a good portion of
the background LMC field.  Total exposure times and average seeing
conditions for each cluster are listed in Table~1.

De-biasing and flat-fielding were performed with standard IRAF
packages, using the calibration frames (bias and sky flat fields)
obtained during the same night of the target observations, or one
night before.  Photometric reduction was carried out with the
DAOPHOT~II/ALLFRAME package (Stetson 1987, 1994). Stellar PSF models
for each frame were obtained by means of a large set of uncrowded and
unsaturated stars. After performing aperture photometry on each frame,
spatial transformations were determined among all the frames of each
cluster, in order to be able to register them and obtain a median
image free from cosmic rays and having high S/N ratio.  This image was
used for star finding, in order to obtain a deeper catalog to use as
input for the simultaneous profile-fitting photometry of all the
frames, made with ALLFRAME.

According to the log file of the service observations, the present
data were obtained under unstable photometric conditions: only part of
the nights were photometric, and some cirrus seemed to be always
present close to the horizon. Therefore, calibration to the Johnson
standard photometric system was performed by comparison of the stars
in common with another set of images of the same clusters, obtained by
our group at the CTIO 4m telescope using the MOSAIC camera, on the
nights Nov. 30th to Dec 1st 1999, which were devoted to a different
project (Gallart et al. 2002).  Those images, although not deep
enough for our purpose, were obtained under photometric conditions,
and were already reduced and calibrated, by means of a set of Landolt
(1992) standard fields observed on the same nights. The $V,R$ cluster
images were obtained on chip\#2 of the MOSAIC camera, on which we
observed around 50 Landolt standards on the color range
$(V-R)=0\div1$. No color terms were considered necessary on the
photometric transformation, and zero-points had typical errors of
$\pm$ 0.004. Aperture corrections for the cluster stars were obtained
from a list of $\simeq$ 70 relatively isolated, bright stars in each
cluster, with standard errors of $\pm 0.002$ mag. Therefore the
total maximum zero-point in the calibration of the CTIO images is
of $\pm$ 0.005 magnitudes. 
The FORS1 data were then calibrated by means of the $\sim 100$
brightest and most isolated stars in common with the CTIO data,
from which we derived a zero point and a color term, with a total
error (including the one quoted above) of $\pm 0.006$.

In the two following papers in this series, we will use both the
morphology and the {\it number} of observed stars in different regions
of the CMDs as compared with theoretical predictions. To this end, it
is crucial to correct the observed counts for completeness.
Completeness corrections were determined by means of artificial-star
experiments. A total of about 30,000 stars were added to the original
frames, with magnitudes and colors consistent with the main branches
of the cluster and field CMD. In order to avoid overcrowding, while
optimizing the CPU time, the artificial stars were added along the
corners of an hexagonal grid, as explained in Zoccali et al. (2000),
in five independent experiments. As usual, photometry of the
artificial frames was performed in a way identical to that of the
original ones.


\section{UNVEILING THE CLUSTER CMDs: SUBTRACTION OF THE LMC FIELD STARS}
\label{sec_cmd}

\begin{figure*}
\centerline{\psfig{figure=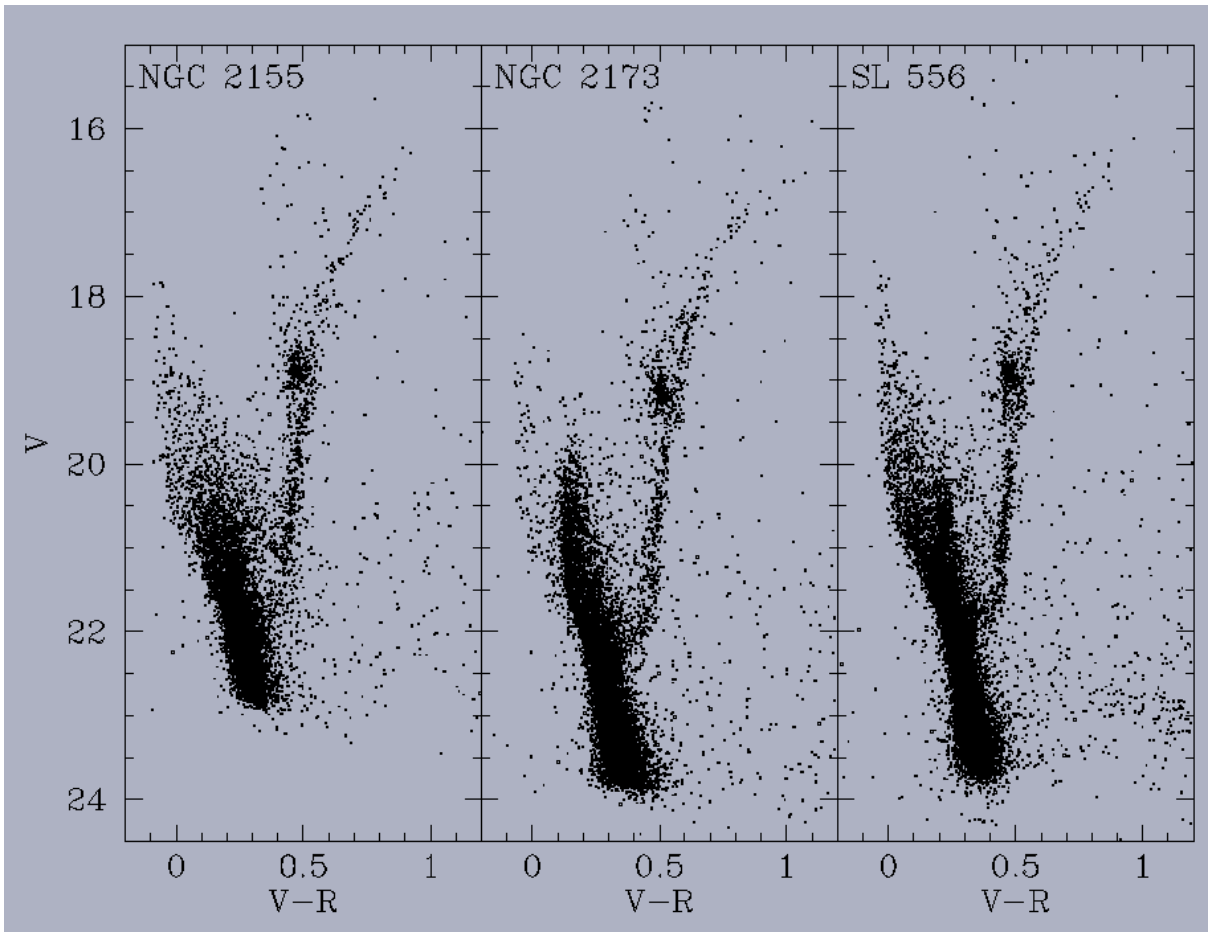,width=16cm,angle=-90}}
\caption{CMDs of NGC~2155, NGC~2173 and SL~556.}
\label{cmds}
\end{figure*}

Figure~\ref{cmds} shows the CMDs of the three LMC clusters NGC~2173,
SL~556 and NGC~2155. All the stars measured with good precision in the
whole FORS1 frames are plotted here: therefore the CMDs include both
cluster and field stars. In order to obtain a cleaner cluster CMD, we
proceeded in the following way.

\begin{figure*}
\centerline{\psfig{figure=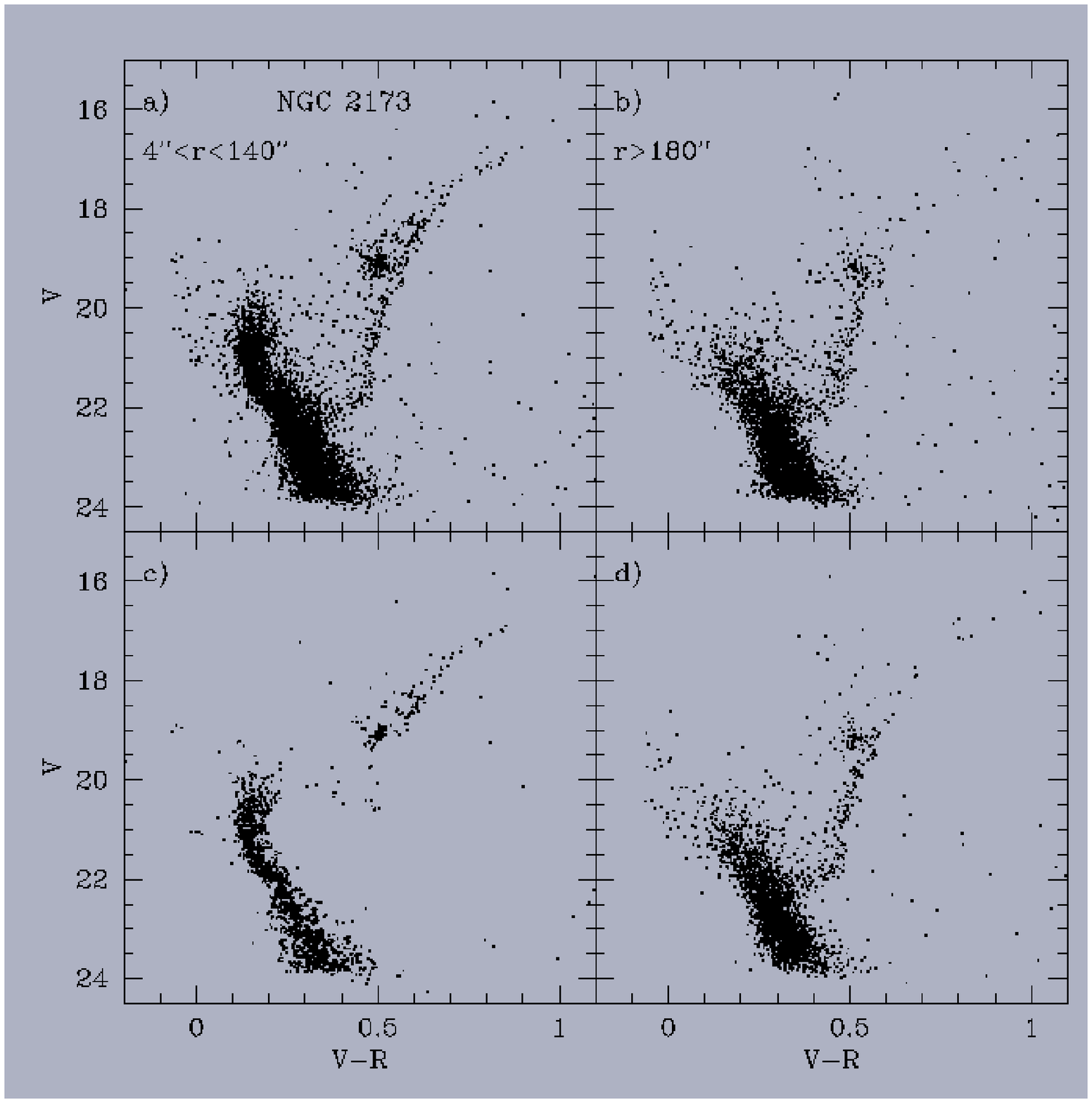,width=16cm}}
\caption{Statistical decontamination from LMC field stars of the
CMD of NGC~2173. {\it a)} Contaminated cluster CMD, as measured in
an annulus around the cluster center; {\it b)} Field CMD, as extracted
from the outer part of the frame; {\it c)} Decontaminated cluster CMD;
{\it d)} Stars subtracted from the CMD of panel {\it a} in order
to obtain {\it c}.}
\label{n2173}
\end{figure*}

\begin{figure*}
\centerline{\psfig{figure=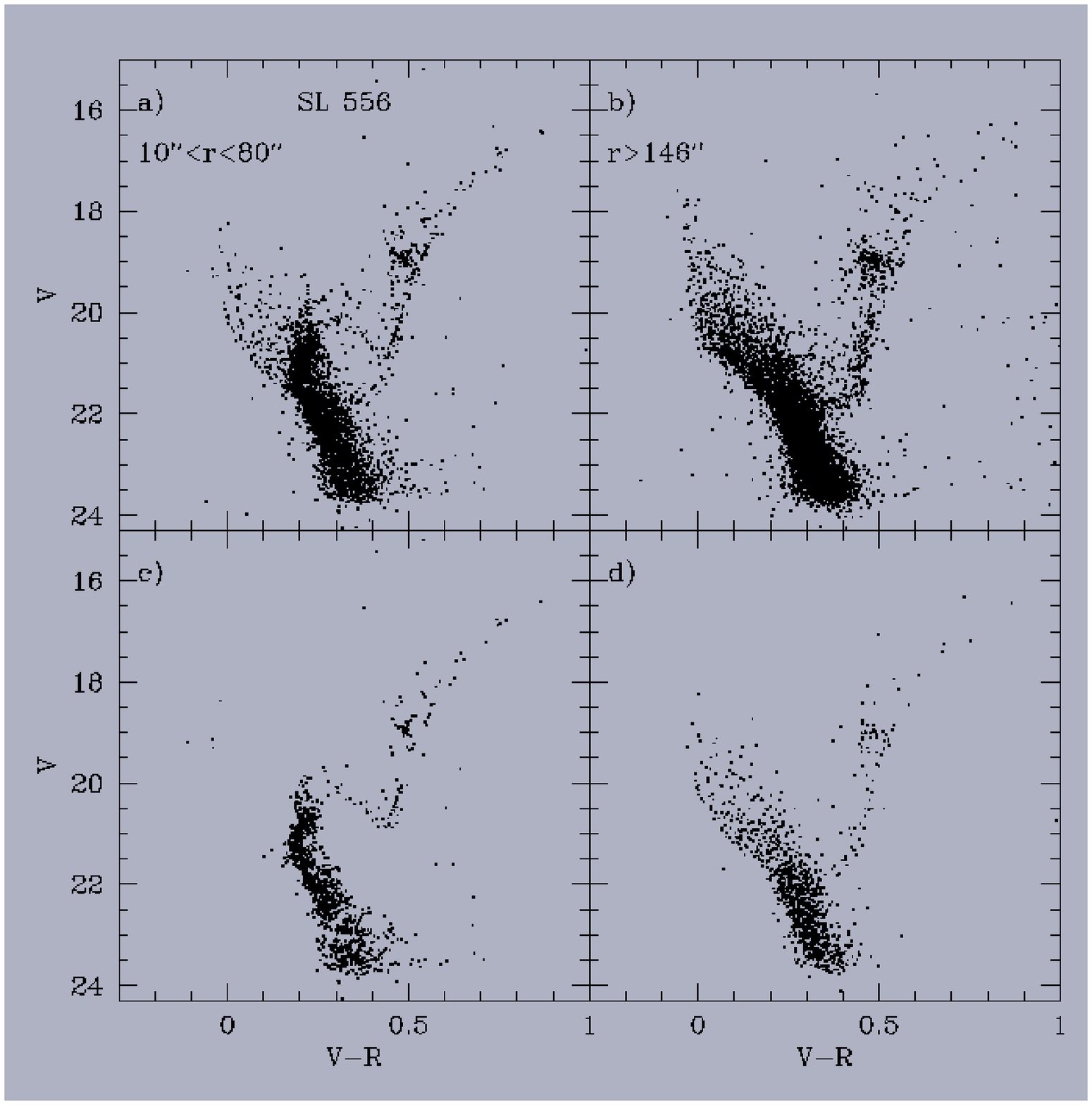,width=16cm}}
\caption{Statistical decontamination from LMC field stars of the
CMD of SL~556. {\it a)} Contaminated cluster CMD, as measured in
an annulus around the cluster center; {\it b)} Field CMD, as extracted
from the outer part of the frame; {\it c)} Decontaminated cluster CMD;
{\it d)} Stars subtracted from the CMD of panel {\it a} in order
to obtain {\it c}.}
\label{sl556}
\end{figure*}

\begin{figure*}
\centerline{\psfig{figure=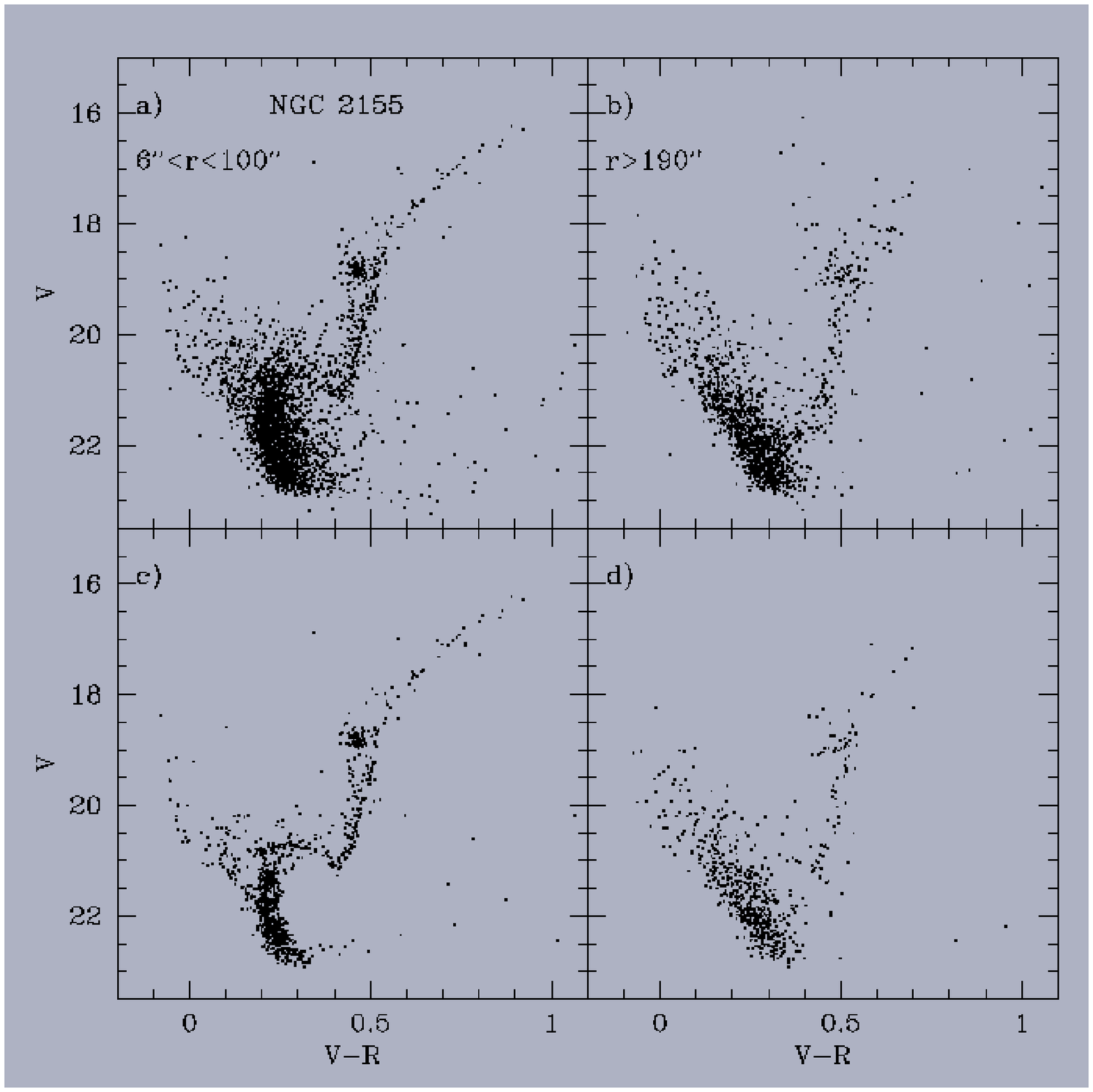,width=16cm}}
\caption{Statistical decontamination from LMC field stars of the
CMD of NGC~2155. {\it a)} Contaminated cluster CMD, as measured in
an annulus around the cluster center; {\it b)} Field CMD, as extracted
from the outer part of the frame; {\it c)} Decontaminated cluster CMD;
{\it d)} Stars subtracted from the CMD of panel {\it a} in order
to obtain {\it c}.}
\label{n2155}
\end{figure*}

For each cluster, an annulus around the center was selected with the
criterion of being large enough to contain most cluster stars, but
small enough to minimize the background contribution. The CMD for the
stars in this annulus are shown in the upper left panel (panel {\it
a}) of Figs.~\ref{n2173} to \ref{n2155}, together with the inner and
outer radii of the selected annuli. The small circular region
containing the very center of the cluster was excluded because stars
in these areas were too crowded, and therefore their photometry was
rather poor.  A CMD typical of LMC field stars was extracted from a
second region, outside a given radius around the cluster center,
selected with the criterion of containing a number of stars large
enough to insure good statistics, but far enough from the cluster not
to contain a significant fraction of cluster stars.  The CMDs
extracted from these ``field'' regions are shown in the upper right
panels ({\it b}) of Figs.~\ref{n2173} to \ref{n2155}, together with
the inner boundary of the region. The outer boundary is set by the
edges of the frame.

For each star in the field CMD (panel {\it b}) we picked up the {\it
closest} (see below) star in the cluster CMD (panel {\it a}), and
decided whether to subtract it or not according to a factor given by
the ratio of the areas of the two regions, multiplied by the ratio of
the completeness factors for the two regions (being less crowded, the
field CMD is more complete than the cluster one). The {\it distance},
$d$, from the field star to each cluster star on the CMD was actually
computed by enhancing the difference in color by a factor of 7 with
respect to the difference in magnitude, i.e.:
\[
d=\sqrt{ [7\times\Delta(V-R)]^2 + \Delta V^2}.
\]
which defines an ellipse with major axis along the magnitudes and minor
axis along the colors.
The reason for this is that given a certain star in the field CMD,
its ``twin'' in the cluster CMD has a color that is better constrained
than the magnitude, because any small difference in either distance,
reddening, or mass between the two would result in a difference in
magnitude larger than that in color.  If there were no stars closer
than a maximum distance of $d=0.42$ magnitudes then no star was
subtracted from the cluster CMD, and the field star was flagged as
unsubtracted.  Both the enhancement factor and the maximum
distance are somewhat arbitrary, and they were chosen after various
experiments, according to two criteria: {\it i.} The CMD of the
subtracted stars (panel {\it d}) should be as similar as possible to
that of the field stars (panel {\it b}) and {\it ii.} the number of
field stars flagged as unsubtracted should be as small as possible.

The result of this procedure (i.e., the CMD in panel {\it c}) is
indeed satisfactory, since in general only the CMD of a single
intermediate-age population is present in this plot. Some younger main
sequence stars are still present in the decontaminated CMD of
NGC~2155, due to the fact that, being the young LMC population
intrinsically clumpy (like any young stellar population in galaxies),
the assumption we make here, that the stars we measure just outside
the cluster boundary are representative of a population identical to
that contaminating the cluster CMD, is true only as a first
approximation.

\section{A COMPARISON BETWEEN THE YONSEI-YALE AND PADOVA STELLAR
EVOLUTIONARY MODELS}

The goal of this project is to compare the observed and theoretical
CMD of intermediate-age, metal-poor clusters, in order to constrain
some of the still uncertain parameters used in stellar evolutionary
models.  In the following two papers of this series, the data
presented here will be independently analyzed by members of the Yale
and the Padova stellar evolution groups, with the aim of calibrating
both sets of models, and also to better identify the effects of
different assumptions on the simulated CMDs.  In view of this, here we
present a comparison between the sets of isochrones published by the
two groups, in the range of ages and metallicity spanned by the
observed clusters, namely 1$\div$3 Gyr and Z=0.004$\div$0.008. We
consider this as a ``first-order'' comparison because we are
considering only the published models, without any fine tuning of the
input parameters, that will be performed in the following two papers.
Because the two grids of models and isochrones were derived
independently, a comparison between them can provide a measure of the
uncertainties in isochrone fitting, as well as the robustness of the
ages derived for LMC star clusters.

\begin{figure*}
\centerline{\psfig{figure=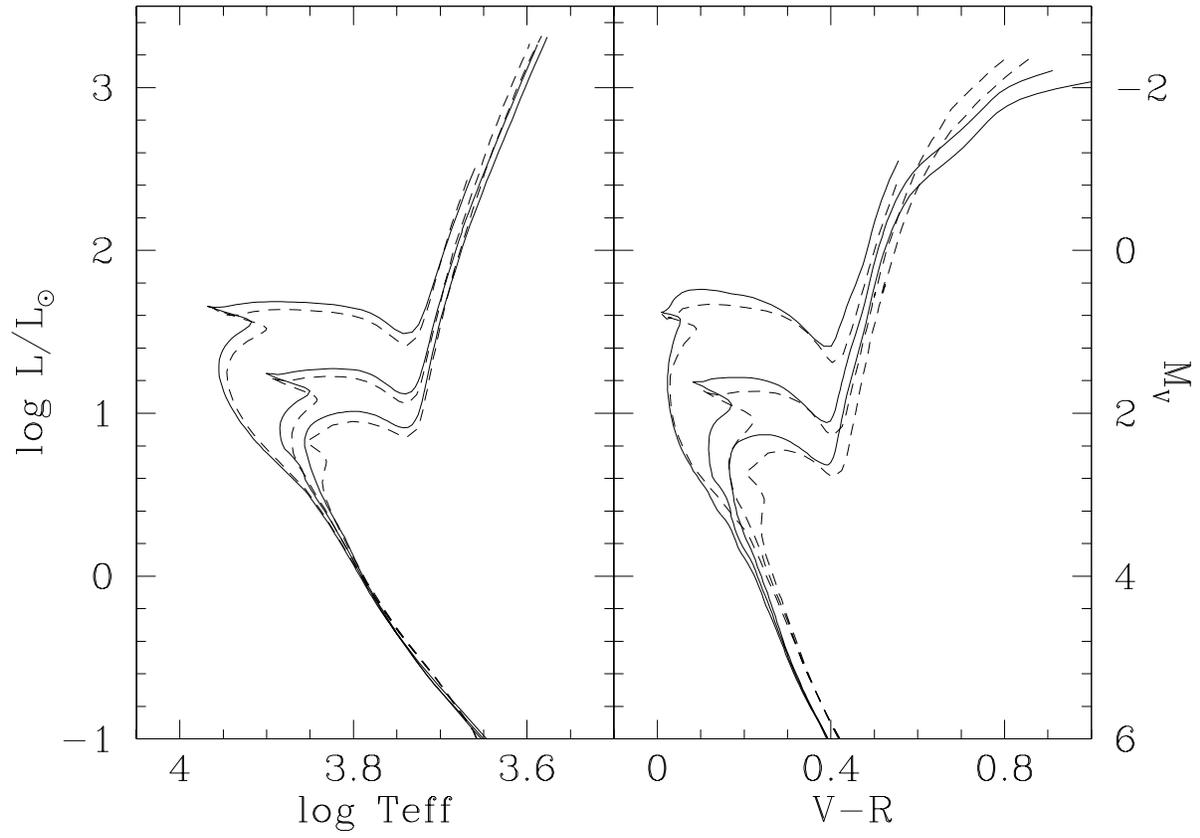,width=16cm,angle=-90}}
\caption{Comparison of the Y$^2$ 
(solid lines) and Padova isochrones (dashed lines) for Z=0.004 and
ages 1, 2 and 3 Gyr. The theoretical plane is shown in the left panel,
while the observacional $[(V-R),V]$ CMD is shown on the right.}
\label{compayalepado}
\end{figure*}

Figure~\ref{compayalepado} shows a set of isochrones of ages 1, 2 and
3 Gyr, and Z=0.004, obtained from Girardi et al. (2000) and Yi et
al. (2001) in the theoretical [log $L/L_{\odot}$,log $T_{\rm eff}$]
and in the observational $[V,(V-R)]$ plane.  The two sets of
isochrones are quite close to one another up to the main sequence
turnoffs in the theoretical plane. Above that, the Y$^2$ isochrones are
slightly brighter than the Padova ones, while the RGB is slightly
cooler.  Another apparent difference is in the 3 Gyr old isochrone,
which is calculated from models with a moderate amount of core
overshooting (gradually increasing with mass in the range between 1.0
and $1.5 M_\odot$, see Girardi et al. 2000) in the Padova set, but
with no overshooting for Y$^2$ isochrones by Yi et al. (2001), who
adopted some overshooting only for isochrones younger than 3 Gyrs. The
differences are more important in the observational plane (right
panel), where the uncertainties in the color transformations add
up. In the following we will discuss in some detail the differences in
the input physics and color transformations between both sets of
models, which may explain the differences apparent in the figures.

In the theoretical log~$L/L_\odot$ vs. log~$T_{eff}$ plane shown in
the left-hand panel of Figure~\ref{compayalepado}, the differences
must be due to differences in the stellar models used in constructing
the isochrones.  Regarding the input physics, both sets of isochrones
make use of the OPAL opacities (Rogers \& Iglesias 1995) for the
stellar interiors, and of the Alexander \& Ferguson (1994) opacities
at low temperatures.  But they differ in the choice of the equation of
state in the interior.  While the Y$^2$ isochrones are based on the
OPAL equation of state (Rogers, Swenson \& Iglesias 1996), the Padova
isochrones use the prescription by Straniero (1988) and the MHD
equation of state (Mihalas, D\"{a}ppen \& Hummer 1988).  Also
different are the energy generation rates and the details of the solar
calibration, which determine the choice of two important input
parameters, the initial helium abundance and the mixing length in the
convective zone.  The mass-luminosity relation for stars depends
significantly on the chosen helium abundance, and the mixing length
parameter in the convection zone of cool stars affects the radius of
cool star models.  

Within the conventional method of overshoot parameterization Padova
isochrones differ from Y$^2$ ones in the adopted constraints to the
overshoot efficiency, as described by Bressan et al. (1981) and
Girardi et al. (2000), and by Yi et al. (2001) and Woo \& Demarque
(2001).  In the Y$^2$ models, the extension of the convective motions
beyond the formal boundary of the convective core defined by the
Schwarzschild criterion is measured from the convective core boundary
outward (i.e. {\it above} the convective boundary).  Its magnitude is
expressed as a fraction of the local pressure scale height $H_{p}$ at
the convection boundary.  In the Y$^2$ models, OS=0.2 (meaning
0.2$H_{p}$) is adopted for younger isochrones ($\le$ 2 Gyr) and OS=0.0
for older ones ($\ge$ 3 Gyr).  In the Padova models, the overshoot
length $\Lambda_{c}$ is measured {\it across} the formal Schwarzschild
convective boundary (i.e. it is measured from inside the convective
core and it straddles the convective boundary).  The overshooting
parameter $\Lambda_{c}$ is so defined that overshoot of $\Lambda_{c}$
= 0.5 (Padova) corresponds to approximately OS = 0.25 $H_{p}$ in the
Yale code. Understanding this difference will be important to
correctly interpret the conclusions on the need of a particular amount
of overshoot in the models, in the next two papers in this series.

Despite these main differences and some minor others in the
construction of the stellar models, it is encouraging to note the
generally good agreement in the log~$L/L_\odot$ vs. log~T$_{eff}$
plane, and in the derived ages.

In the CMD, shown on the right-hand side in
Figure~\ref{compayalepado}, the differences between the Y$^2$ and
Padova isochrones are more pronounced.  This emphasizes the still
large uncertainties in the color transformation, used to translate the
isochrones from the theoretical log~$L/L_\odot$ vs. log~$T_{eff}$
plane into the observational $M_V$ vs. $(V-R)$ plane. Bertelli et
al. (1994) and Girardi et al. (2000) isochrones are based on the
Kurucz (1992) library of stellar spectral energy distributions,
convolved with filter spectral responses (extended at high temperature
with black-body spectra and at low temperature with empirical M stars
spectra as described in those paper).  

The Y$^2$ isochrones used here are instead
based on the semi-empirical Lejeune, Cuisinier \& Buser
(1998) color transformations (BaSeL colors). The spectral libraries
assembled by Lejeune et al. are based on the Kurucz (1995) library,
extended to low temperatures based on empirical data and
low-temperature stellar models: for M giants, spectra from Fluks et
al.  (1994) and Bessell et al.  (1989,1991), and for M dwarfs,
synthetic spectra from Allard \& Hauschildt (1995), were used to
correct spectral distributions with semi-empirical calibrations.
There is also a version of the Y$^2$ set of isochrones using the color
transformations by Green et al. (1987), but those transformed with the
Lejeune et al. (1998) tables have been selected as those better
reproducing the observed CMDs.

The effects of the different color transformations are apparent in the
[$V,(V-R)$] plane shown in the rigth panel. The Padova isochrones
(Girardi et al. 2000) are slightly redder than the Y$^2$ ones along
the unevolved main sequence and in the RGB, an effect that must be
attributed to the color transformation alone since these parts of the
isochrones match closely in the theoretical plane.  The differences
in the sub giant branch (SGB) region, mainly consisting in the Y$^2$
isochrones being brighter, are a combination of the differences in
luminosity already noted in the theoretical plane, and the effects of
the color transformation. In any case, the uncertainty in color is
within the errors of the Hipparcos subdwarf data.


\section{ISOCHRONE FITTING TO THE CLUSTER CMDs}
\label{sec_age}

\begin{figure*}
\centerline{\psfig{figure=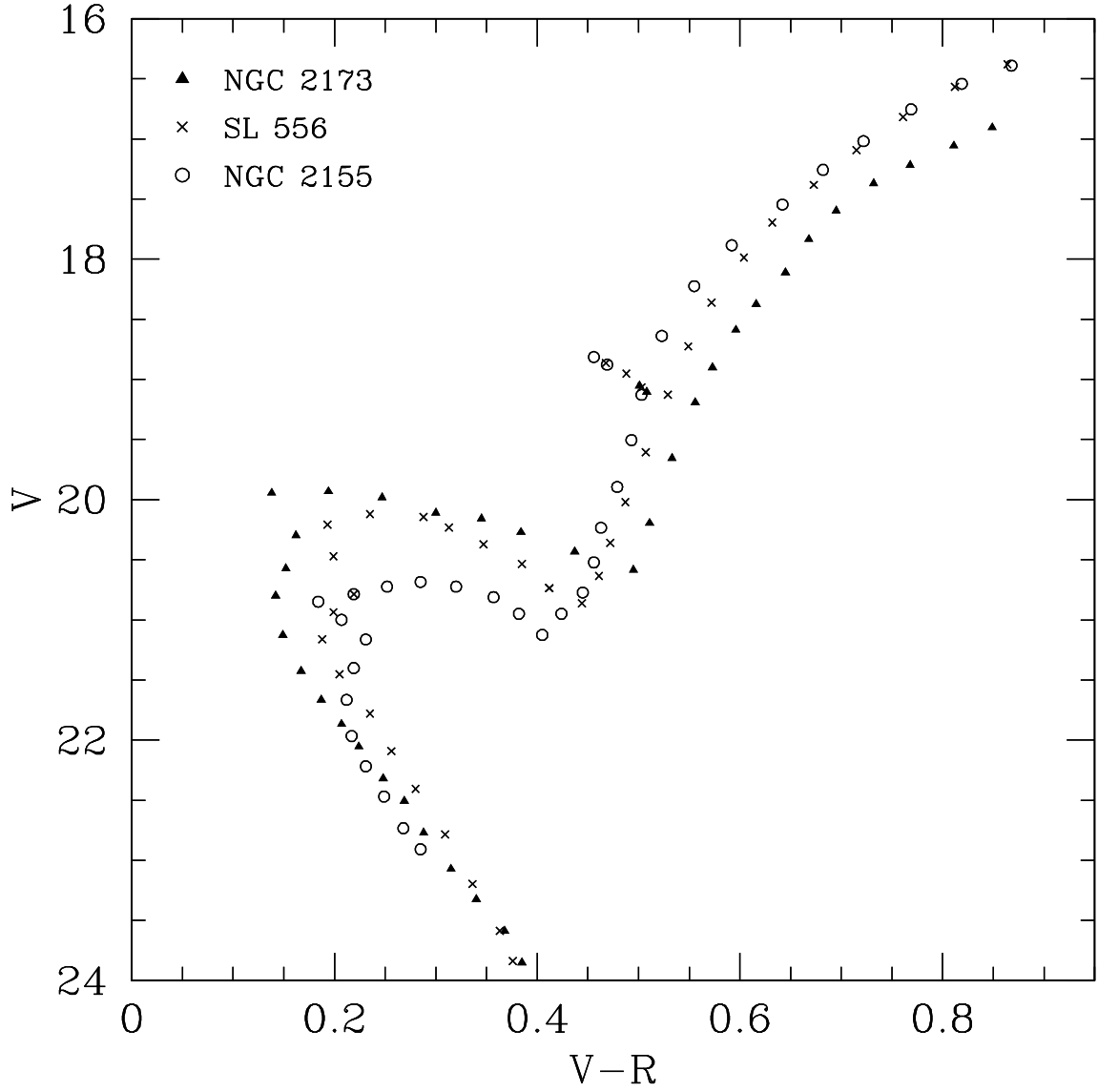,width=16cm,angle=0}}
\caption{Comparison of the CMD loci for the three clusters.
Note the sequence of ages in the main sequence turnoff.}
\label{compaclus}
\end{figure*}

Figure~\ref{compaclus} displays a comparison among the fiducial lines
of the three clusters. The cluster loci have been superimposed with no
correction for possible differences in reddening or distance
modulus. Note that, due to the presence of binary stars (see
discussion in Paper II), drawing a fiducial sequence in the SGB region
just above the main sequence turnoff is not straightforward,
especially for SL~556. Nevertheless, it is still evident from
Figure~\ref{compaclus} that the clusters show a trend in the $V$
magnitude of the main sequence turnoff, and a (opposite) trend in the
magnitude of the red-clump. The main sequence trend is clearly one in
age, while the variation of the red-clump luminosity is consistent
with the prediction by Girardi \& Salaris (2001) about the luminosity
evolution of the red-clump in the age interval 1-3 Gyr.

\subsection{NGC~2173}

\begin{figure*}
\centerline{\psfig{figure=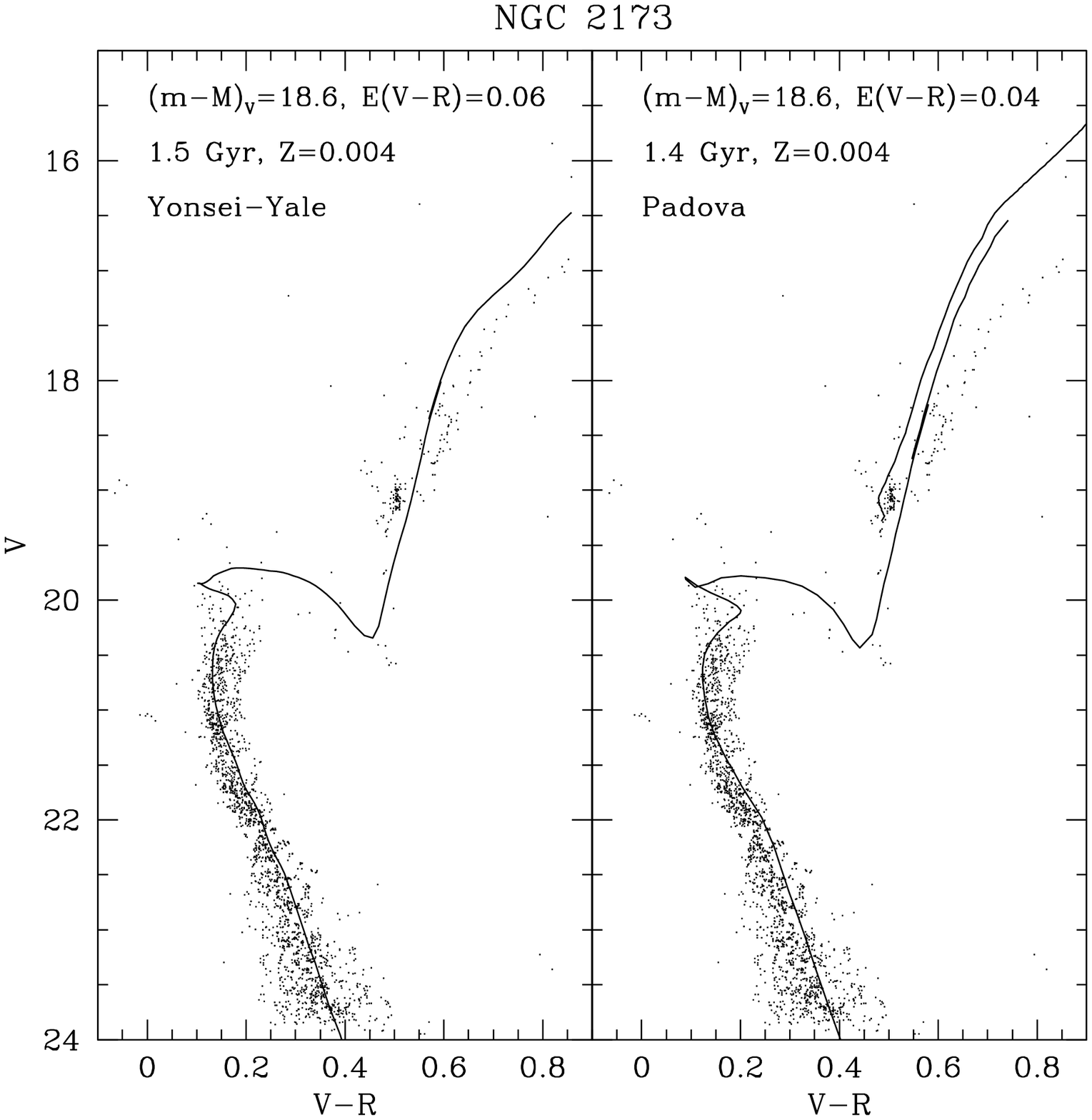,width=16cm,angle=0}}
\caption{Comparison between the CMD of NGC~2173 and theoretical
models from the Yale group (left) and the Padova group (right).
The adopted cluster parameters are listed in the figure labels.}
\label{n2173_iso}
\end{figure*}

Figure~\ref{n2173_iso} shows the comparison between the CMD of
NGC~2173 and the Y$^2$ (left) and Padova (right) theoretical
isochrones. The same apparent distance modulus $(m-M)_V=18.6$ and
metallicity Z=0.004, have been used in both cases, while due to the
differences in the color transformations shown in
Fig.~\ref{compayalepado}, slightly different reddenings and ages are
required for a good match with the two models. Both models reproduce
quite well the main sequence, turnoff and SGB region, while they both
fail to reproduce the color location of the RGB. A small change in
slope would be needed to bring the theoretical Y$^2$ RGB on top of the
observed one, while a substantially different slope would be required to
bring the Padova model into agreement with the data. In both cases a
higher metallicity would improve the RGB fit, while the fit to the
rest of the CMD remains similarly good. We adopt Z=0.004 for
consistency with the two following papers in this series. The
magnitude of the red-clump of the Padova models matches very well the
observed one, while the color is slightly too blue, as all the
AGB. Our crude estimate of the cluster age is therefore $\sim$ 1.5
Gyr.

\subsection{SL~556}

\begin{figure*}
\centerline{\psfig{figure=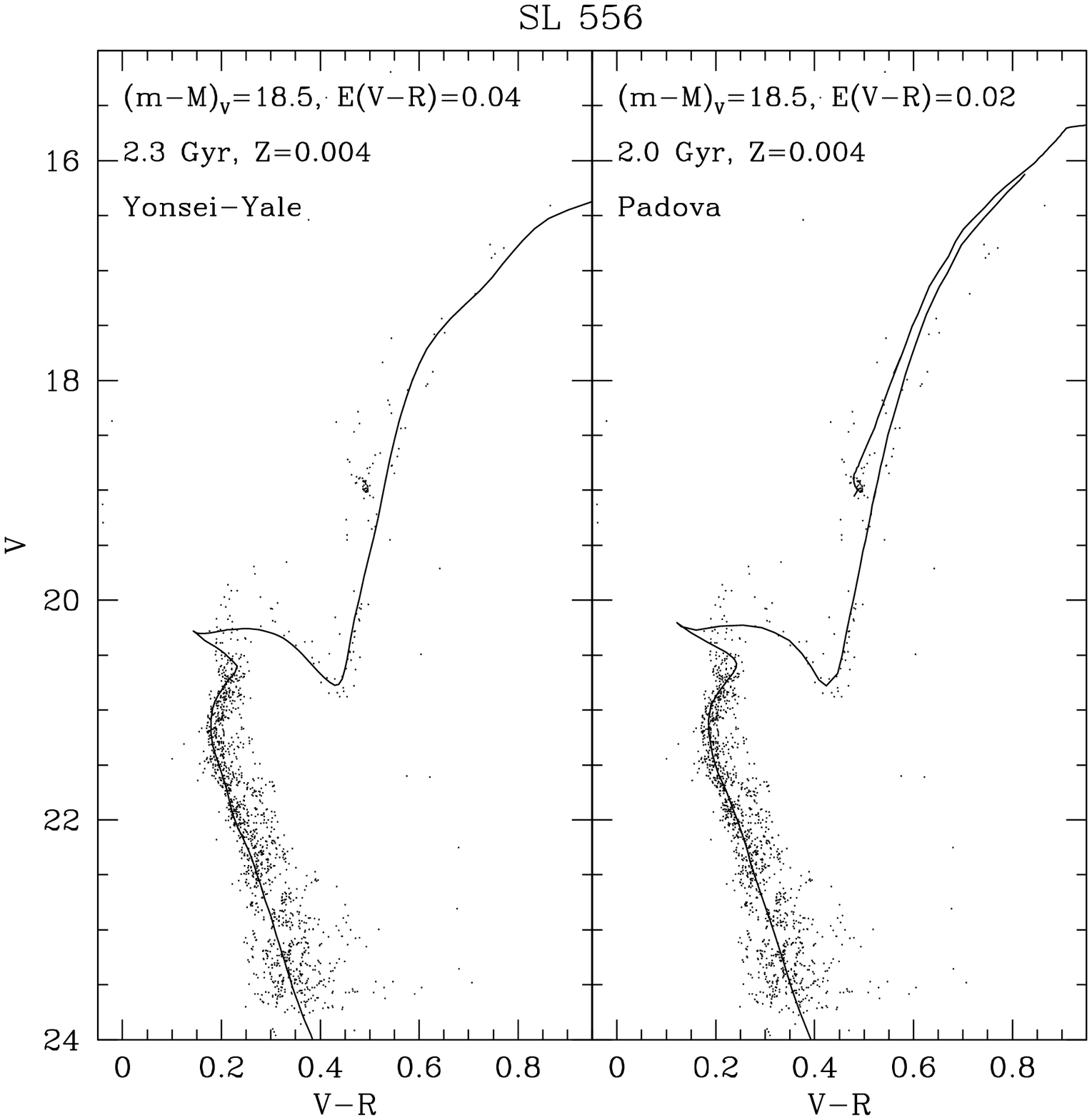,width=16cm,angle=0}}
\caption{Comparison between the CMD of SL~556 and theoretical
models from the Yale group (left) and the Padova group (right).
The adopted cluster parameters are listed in the figure labels.}
\label{sl556_iso}
\end{figure*}

Figure~\ref{sl556_iso} shows the comparison between the CMD of SL~556
and the theoretical isochrones from Y$^2$ (left) and Padova (right). A
distance modulus of $(m-M)_V=18.5$ and a metallicity of Z=0.004 has
been adopted in both cases, while slightly different reddening and age
were used for the two models. The shape of the SGB region of this
cluster is rather peculiar, and apparently quite different from
theoretical predictions. As it will be shown in the next papers of
this series, this shape can be explained by the presence of a fraction
of binaries, also evident from the fact that there seems to be a
second sequence, on the right, and parallel, to the cluster main
sequence.  Excluding this SGB region just above the turnoff, both
models reproduce well the observed CMD. The Y$^2$ models also predict
a correct RGB slope all the way up to the tip, while the Padova ones
are steeper than the observed RGB, for magnitudes brighter than the
red-clump. Also in this case, the fit would be improved by adopting a
higher metallicity, but then also a negative reddening would be
required.

\subsection{NGC~2155}

\begin{figure*}
\centerline{\psfig{figure=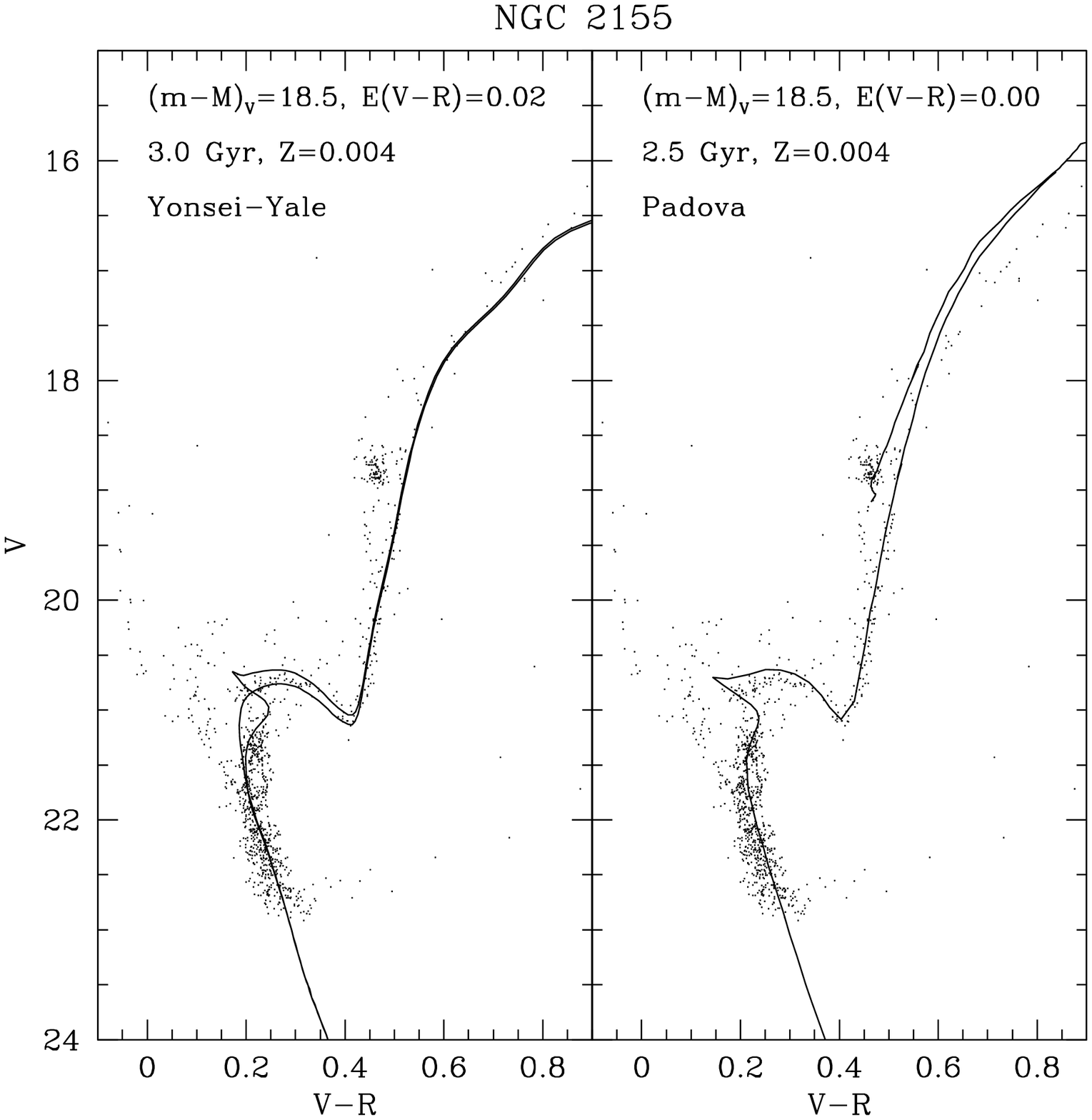,width=16cm,angle=0}}
\caption{Comparison between the CMD of NGC~2155 and theoretical
models from the Yale group (left) and the Padova group (right).
The adopted cluster parameters are listed in the figure labels.}
\label{n2155_iso}
\end{figure*}

Figure~\ref{n2155_iso} shows the CMD of NGC~2155 compared with Y$^2$
(left) and Padova theoretical isochrones (right). Also for this
cluster the same apparent distance modulus $(m-M)_V=18.5$ and
metallicity Z=0.004, have been used in both cases, while slightly
different reddenings and ages were required to achieve a good fit with
the two models. Two isochrones are shown in the left panel, with or
without core overshooting, the difference being only evident in the
shape of the turnoff and SGB; some intermediate value of overshooting
seems to be needed for a better fit of the observed points (see
discussion in Paper II). The RGB is well reproduced by the $Y^2$ model
all the way up to the tip, while the Padova one, due to the use of
different transformations, starts to deviate from the observed points
towards brighter magnitudes. A higher metallicity would improve the
fit of the RGB slope, but then the other sequences would be too red in
the model. The red-clump, only available in the Padova isochrones,
perfectly matches the observed one, while the theoretical AGB is of
course too blue, as the upper RGB. Using these models, our best
estimate of the cluster age in this preliminary analysis is therefore
between 2.5 and 3 Gyr.


\section{SUMMARY}
\label{summ}

This preliminary investigation on three intermediate age LMC clusters,
namely NGC 2173, SL 556 and NGC 2155 is the starting point for a very
detailed analysis aiming at a test on input physics for stellar
models.  In principle there is the opportunity to check the efficiency
of convective core overshoot for star masses between 1.1 and 1.5
$M_{\odot}$ by comparison of these CMDs with synthetic CMDs and
isochrones.  The isochrone fitting to the CMDs of the clusters gives
some information on their age and metallicity, but we must take into
account that there are also uncertainties in the LMC distance modulus
and interstellar reddening.  Only a more refined analysis, considering
also the star distribution in several regions of the CMD, will give
reliable results.  In the following two papers of this series, the
synthetic CMD technique will be used independently by members of the
Padova and Yale groups to determine the cluster characteristics,
taking into account the uncertainties in the observations and in
stellar evolutionary models, and to give some feed-back into the
stellar evolutionary models themselves.


\acknowledgements 

Our data was collected as part of an ESO Service Mode
run. C.G. acknowledges partial support from chilean CONICYT through
FONDECYT grant number 1990638. This research has also been supported
in part by NASA grant NAG5-8406 (P.D.).



\newpage
\begin{figure*}
\caption{SL~556 as imaged with FORS1@VLT-UT1; North is up and East
is to the left.}
\label{frame}
\end{figure*}

\newpage

\newpage

\newpage

\newpage

\newpage

\newpage

\newpage

\newpage

\newpage

\vfil
\clearpage

\begin{deluxetable}{ccccccc}
\tablewidth{35pc}
\tablenum{1}
\tablecaption{Log of observations}
\tablehead{
\colhead{Object}  &
\colhead{Date}    &
\colhead{RA}      &
\colhead{DEC}     &
\colhead{Filter}  &
\colhead{Exptime} &
\colhead{Seeing}  }
\startdata
NGC~2155 & 11 Jan 2000 & 05:58:32.4 & -65:28:38.4 & V & 3768s & $\sim$0.9 \\
   ``    &     ``      &     ``     &      ``     & R & 2235s & $\sim$0.9 \\
NGC~2173 & 12 Jan 2000 & 05:57:39.7 & -72:58:41.4 & V &  928s & $\sim$0.6 \\
   ``    &     ``      &     ``     &      ``     & R &  855s & $\sim$0.6 \\
SL~556	 & 13 Dec 1999 & 05:32:25.0 & -64:44:11.0 & V & 1856s & $\sim$0.6 \\
   ``    &     ``      &     ``     &      ``     & R & 1710s & $\sim$0.6 \\
\tableline
\enddata
\end{deluxetable}


\begin{references}

\reference{} Alexander, D.R. \& Ferguson, J.W. 1994, ApJ, 437, 879
\reference{} Allard, F. \& Hauschildt, P.H., 1995, ApJ, 445, 433
         \& Schweitzer,A., 2001, ApJ, 556, 357
\reference{} Aparicio, A., Bertelli, G., Chiosi, C., \& Garcia-Pelayo, J.M., 1990,
        A\&A, 240, 262
\reference{} Barbaro, G. \& Pigatto, L., 1984, A\&A, 136, 355
\reference{} Barmina, R., Girardi, L., \& Chiosi, C., 2002, A\&A, 385, 847
\reference{} Benedict, G.F. et al. 2002. AJ, 110, 212
\reference{} Bertelli, G., Bressan, A., Chiosi, C., Fagotto, F. \& Nasi, E.
	1994, A\&AS, 106, 275
\reference{} Bertelli, G., et al. 2002, in preparation (Paper III)
\reference{} Bessell M.S., Brett J.M., Scholz M. \& Wood P.R. 1989, A\&AS 77, 1 
\reference{} Bessell M.S., Brett J.M., Scholz M. \& Wood P.R. 1991, A\&AS 89, 335

\reference{} Bica, E., Dottori, H., \& Pastoriza, M. 1986, A\&A, 156, 261
	A\&A, 293, 710
\reference{} Bressan, A., Bertelli, G., \& Chiosi, C., 1981, A\&A, 102, 25
\reference{} Brocato, E. \& Castellani, V. 1988, A\&A, 203, 293
\reference{} Brocato, E., Castellani, V. \& Piersimoni, A.M. 1994, A\&A, 290, 59
\reference{} Carney, B. W. 1996, PASP, 108, 900
\reference{} Carraro, G., Bertelli, G., Bressan, A., \& Chiosi, C., 1993,
        A\&AS, 101, 381
\reference{} Chiosi, C., Bertelli, G., Meylan, G. \& Ortolani, S. 1989, A\&AS, 78, 89
\reference{} Chiosi, C., Bertelli, G., Meylan, G. \& Ortolani, S. 1989, A\&A, 219, 167
\reference{} Chiosi, C. \& Pigatto, L., 1986, ApJ, 308, 1
\reference{} Chiosi, C., Vallenari, A., Bressan, A., Deng, L., \& Ortolani, S. 1995,
\reference{} Cohen, J. 1982, ApJ, 258, 143
\reference{} Corsi, C.E., Buonanno, R., Fusi Pecci, R., Ferraro, F.R., Testa,
	V., \& Greggio, L. 1994, MNRAS, 271, 385
\reference{} Demarque, P., Sarajedini, A., \& Guo, X.-J. 1994, ApJ, 426, 165
\reference{} Dinescu, D.I., Demarque, P., Guenther, D.B., \& Pinsonneault, M.H. 1995,
	AJ 109, 2090
\reference{} Fagotto, F., Bressan, A., Bertelli, G., \& Chiosi, C., 1994, A\&AS,
	105, 39
\reference{} Fluks, M.A., Plez, B., The, P.S., de Winter, D., Westerlund, B. E. \& Steenman, H. C. 1994, A\&AS, 105, 311 
\reference{} Gallart et al. 2002, in preparation.
\reference{} Geisler, D., Bica, W., Dottori, H., Claria, J.J., Piatti, A.E., \&
	Santos, J.F.Jr. 1997, AJ, 114, 1920
\reference{} Girardi, L., Bressan, A., Bertelli, G., \& Chiosi, C. 2000, A\&AS, 141, 371
\reference{} Girardi, L. \& Salaris, M. 2001, MNRAS, 323, 109
\reference{} Green, E. M., Demarque, P., \& King, C. R. 1987, The Revised Yale Isochrones and Luminosity Functions (New Haven: Yale Univ. Obs.)

\reference{} Greggio, L., \& Renzini, A. 1990, ApJ, 364, 35
\reference{} Heap, S.R., et al. 1998, ApJ, 492, L131
\reference{} Horch, E., Demarque, P., \& Pinsonneault, M., 1992, ApJ, 387, 372
\reference{} Keller, S.C., Da Costa, G.S., \& Bessell, M.S., 2001, AJ, 121, 905
\reference{} Kozhurina-Platais, V., Demarque, P., Platais, I., Orosz, J.A.,
	\& Barnes, S. 1997, AJ, 113, 1045
\reference{} Kurucz, R. 1992, in The Stellar Population in Galaxies, IAU149, eds.
\reference{} Kurucz, R. 1995, private communication
	B. Barbuy \& A. Renzini (Dordrecht: Kluwer), 225
\reference{} Lattanzio, J.C., Vallenari, A., Bertelli, G., \& Chiosi, C. 1991,
	A\&A, 250, 340
\reference{} Landolt, A.U.  1992, AJ, 104, 372
\reference{} Lattanzio, J. C., Vallenari, A., Bertelli, G. \& Chiosi, C. 1991, A\&A, 250, 340
\reference{} Lejeune, Th., Cuisinier, F. \& Buser, R. 1998, A\&A, 130, 65
\reference{} Maeder, A. \& Mermilliod, J.C., 1981, A\&A, 93, 136
\reference{} Mateo, M., \& Hodge, P. 1986, ApJS, 60, 893
\reference{} Mazzei, P. \& Pigatto, L. 1989, A\&A, 213, L1
\reference{} Mermilliod, J.C. \& Maeder, A. 1986, A\&A, 158, 45
\reference{} Meynet, G., Mermilliod, J.-C. \& Maeder, A. 1993, A\&AS, 98, 477
\reference{} Mighell, K.J., Sarajedini, A. \& French, R.S. 1998, ApJ, 116, 2395
\reference{} Mihalas, D., D\"{a}ppen, W. \& Hummer, D.G. 1988, ApJ, 331, 815
\reference{} Mould, J.R., Da Costa, G.S., \& Wieland F.P. 1986, ApJ, 309, 39
\reference{} Olszewski, E.W., Schommer, R.A., Suntzeff, N.B., \& Harris, U.G. 1991,
	AJ, 101, 515

\reference{} Pont, F., Zinn, R., Gallart, C., Winnick, R. \& Hardy, E. 2002, in
	preparation
\reference{} Rich, R.M., Shara, M.M., \& Zurek, D. 2001, AJ, 122, 842
\reference{} Rogers, F.J. \& Iglesias, C.A. 1995, in Astrophysical Applications of
	Powerful New Databases, ed. S.J. Adelman \& W.L. Wiese (San Francisco:
	ASP), 78
\reference{} Rogers, F.J., Swenson, F.J. \& Iglesias, C.A. 1996, ApJ, 456, 902
\reference{} Rosvick, J.M. \& VandenBerg, D.A. 1998, AJ, 115, 1516
\reference{} Salaris, M., Chieffi, A., \& Straniero, O. 1993, ApJ, 414, 580
\reference{} Sarajedini, A. 1998, AJ, 116, 738
\reference{} Stetson, P.\ B. 1987, PASP 99, 191
\reference{} Stetson, P.\ B. 1994, PASP 106, 250
\reference{} Stothers, R.B. \& Chin, C.-W. 1992, \apj, 390, 136
\reference{} Straniero, O., 1988, A\&AS, 76, 157
\reference{} Testa, V., Ferraro, F.R., Chieffi, A., Straniero, O., Limongi, M. \& Fusi Pecci, F. 1999, AJ, 118, 2839
\reference{} Vallenari, A., Chiosi, C., Bertelli, G., Meylan, G., \& Ortolani, S.
        1991, A\&AS, 87, 517
\reference{} Vallenari, A., Aparicio, A., Fagotto, F., \& Chiosi, C. 1994, A\&A,
	284, 424
\reference{} Woo, J.H. \& Demarque, P., 2001, AJ, 122, 1602
\reference{} Woo J.H., et al. 2002, AJ, submitted (Paper II)
\reference{} Yi, S., Demarque, P. \& Oemler A.Jr. 1998, ApJ, 492, 480
\reference{} Yi, S., Brown, T.M., Heap, S., Hubeny, I., Landsman, W., Lanz, T., \&
	Sweigart, A. 2000, ApJ, 533, 670
\reference{} Yi, S., Demarque, P., Kim, Y.-C., Lee, Y.-W., Ree, C.H.,
	Lejeune, Th. \& Barnes, S. 2001, ApJS, 136, 417 (Y$^2$)
\reference{} Zoccali M., Cassisi S., Frogel J.A., Gould A., Ortolani S., Renzini A.,
	Rich R.M., \& Stephens A.W., 2000, ApJ, 530, 418

\end{references}
\end{document}